\documentclass[twocolumn,showpacs,preprintnumbers,amsmath,amssymb,mathrsfs]{revtex4}

\usepackage{graphicx}
\usepackage{dcolumn}
\usepackage{bm}

\newcommand\ba{\begin{align}}
\newcommand\ea{\end{align}}
\newcommand\be{\begin{equation}}
\newcommand\ee{\end{equation}}
\newcommand\bs{\begin{subequations}}
\newcommand\es{\end{subequations}}
\newcommand\nn{\nonumber}
\newcommand\bfl{\begin{flushleft}}
\newcommand\efl{\end{flushleft}}
\newcommand\bsp{\begin{split}}
\newcommand\easp{\end{split}}

\newcommand\ri{\right}
\renewcommand\le{\left}

\newcommand{\sub}[1]{\mbox{\tiny{#1}}}

\renewcommand\a{\alpha}

\renewcommand\c{\psi}
\renewcommand\d{\delta}

\newcommand\f{\phi}

\newcommand\g{\gamma}

\renewcommand\l{\lambda}

\newcommand\p{\pi}

\renewcommand\r{\rho}

\renewcommand\t{\tau}

\newcommand\y{\eta}

\newcommand\pd{\partial}
\newcommand\mc{\mathcal}

\newcommand\mbs{\boldsymbol}
\newcommand\mbf{\mathbf}

\begin{document}
\bibliographystyle{unsrt}



\title{Taylor dispersion with absorbing boundaries: A Stochastic Approach}
\author{Rudro R. Biswas}\email{rrbiswas@physics.harvard.edu}\affiliation{Department of Physics, Harvard University, Cambridge, MA 02138, USA}\altaffiliation[Also at ]{Schlumberger-Doll Research, Ridgefield, CT 06877, USA }
\author{Pabitra N. Sen}
\affiliation{Schlumberger-Doll Research, Ridgefield, CT 06877-4108, USA }

\begin{abstract}
We describe how to solve the problem of Taylor dispersion in the presence of absorbing boundaries using an exact stochastic formulation. In addition to providing a clear stochastic picture of Taylor dispersion, our method leads to closed-form expressions for all the moments of the convective displacement of the dispersing particles in terms of the transverse diffusion eigenmodes. We also find that the cumulants grow asymptotically linearly with time, ensuring a Gaussian distribution in the long-time limit. As a demonstration of the technique, the first two longitudinal cumulants (yielding respectively the effective velocity and the Taylor diffusion constant) as well as the skewness (a measure of the deviation from normality) are calculated for fluid flow in the parallel plate geometry. We find that the effective velocity and the skewness (which is negative in this case) are enhanced while Taylor dispersion is suppressed due to absorption at the boundary. 
\end{abstract}


\maketitle

In a laminarly flowing fluid with flow velocity varying perpendicular to the flow direction, suspended particles will jump randomly across fast and slow streamlines due to transverse diffusion. This fluctuating convectional velocity arising from a random sampling of various streamlines disperses these particles along the direction of flow in a well-studied process known as Taylor dispersion\cite{taylor53,taylor54, aris56, degennes83}. The particles drift along the flow with an average velocity $v_e$ and disperse {\em linearly} with time $t$ along the flow, i.~e.\ the r.m.s displacement in that direction is $\sqrt{(D_{\sub{taylor}}+D) t}$ where the convection-induced Taylor dispersion coefficient $ D_{\sub{taylor}}$ is found to scale as $\ell^2 v_e^2/ D$, $D$ being the molecular diffusion coefficient and $\ell$ being the transverse dimension. Taylor dispersion has turned out to be important both fundamentally in hydrodynamics and in its applications to diverse fields ranging from biological perfusion, chemical reactors and lab-on-chips to soil remediation and oil recovery\cite{saffman59,ajdari06,bontoux06,saffman60,degennes83,scheven02,lowe96,Pfannkuch63}.

In spite of having been around for many decades, Taylor dispersion in the presence of absorption, even in simple geometries (such as Poiseuille flow in tubes), has proven to be a mathematically challenging problem\cite{broeck83, broeck90, gill67, gill73, lungu82}. As Taylor dispersion under Poiseuille flow in tubes is now routinely used to measure molecular diffusion coefficients\cite{cussler97}, it is important to compute the errors due to absorption at the walls. The same holds true for other applications outlined in the previous paragraph. The methods used originally\cite{taylor53,taylor54,aris56} worked only in the absence of absorption. This was pointed out in the most `modern' treatment available --- by Gill, Lungu and Moffatt\cite{gill73,lungu82} dating back to over 40 years ago. They showed that  the effective velocity is enhanced  while Taylor dispersion is suppressed due to absorption at the boundary. However, even in their scheme, obtaining higher than second moments and thus estimating {\em deviation from Gaussianity} of the particle concentration profile along the direction of flow, becomes prohibitively tedious and hence was not addressed.

Our approach goes a step further than this cornerstone work. We provide the correct stochastic picture of the underlying process. When decay is present, the usual probabilistic techniques\cite{broeck83, broeck90} fail. We formulate, for the first time, the correct technique --- ultimately providing a visually appealing and computationally simple stochastic method to find all moments/cumulants of the longitudinal (i.\ e.\ along the convective flow direction) displacement for uniform (linear) laminar flow. We find that the cumulants grow asymptotically linearly with time ensuring a Gaussian distribution in the long-time limit, as is characteristic of drifting random walkers on a lattice\cite{haldane40}.

Consider diffusing particles being carried along by a fluid flowing unidirectionally in the $x$-direction through a uniform cross-section $A$. The particle position in the transverse section $A$ will be denoted by $\vec{y}$, while the complete position will be specified by $\mbs{x}\equiv (x, \vec{y})$. The concentration of the particles, $N(\mbs{x}, t)$, evolves according to the convective diffusion equation inside the fluid volume $V$:
\be\label{eq:N_original}
\frac{\pd N(\mbs{x},t)}{\pd t} = D\,\nabla_{\mbs{x}}^2 N(\mbs{x},t) - \mbs{v}(\mbs{x}).\mbs{\nabla}N(\mbs{x},t) \quad \mbs{x}\in V
\ee
where $D$ is the molecular diffusion constant of the particles in the still fluid and $\mbs{v}(\mbs{x}) \equiv v(\vec{y})\mbs{\hat{x}}$ is the $x$-{\em independent} convective velocity field of the fluid. The presence of absorbing walls is taken into account by the boundary condition
\be
\label{eq:bc_original} D\,\mbs{\hat{e}.\nabla_{x}} N(\mbs{x},t)+\r N(\mbs{x},t)=0, \quad \mbs{x}\in\pd V 
\ee
where $\r$ is a measure of the surface absorption rate that varies from zero (reflecting walls) to infinity (perfect absorption), while $\hat{\mbf{e}}(\mbs{x})$ is the normal to the surface $\pd V$ at $\mbs{x}$. Since the cross-section is uniform, $\hat{\mbf{e}}(\mbs{x})\mbs{.\hat{x}}=0$. This enables us to replace $\mbs{\hat{e}.\nabla_{x}}$ by $\mbs{\hat{e}.\nabla}_{y}$ in \eqref{eq:bc_original}.

By integrating over $x$ in \eqref{eq:N_original} and \eqref{eq:bc_original}, a diffusion equation for $n(\vec{y}, t)=\int dx\,N(\mbs{x},t)$ -- the concentration of particles in the transverse section $A$ -- is obtained with absorbing boundary conditions (when $N, \pd_x N\rightarrow 0$ as \mbox{$x\rightarrow \pm \infty$}):
\bs\label{eq:n_original}
\ba
\frac{\pd n(\vec{y}, t)}{\pd t} &= D\;\nabla_{y}^2\; n(\vec{y}, t) \quad &\vec{y}\in A\\ 
D\;\mbs{\hat{e}.\nabla}_{y}\; n(\vec{y}, t) &= -\r\; n(\vec{y}, t)\quad &\vec{y}\in \pd A
\end{align}
\es
This yields the $\vec{y}$-statistics of the dispersing particles that we will use in the formalism  below involving the equations \eqref{eq:basic} and \eqref{eq:propagator-conc}.

We now proceed to find the moments of $x(t)$. We shall not show explicitly the noise that is responsible for the usual molecular diffusion along the $x$ direction and concentrate on the dominant convective noise\footnote{This type of process is common in many areas of physics like spectral diffusion\cite{Kubo91}. In line-shape problems of NMR/ESR, $x\equiv i\phi = \ln M$ would be the `phase', $\vec{y}$ the position of the diffusing particles while $v(\vec{y})$ would be the local Larmor frequency $\omega(\vec{y})$.} encoded in the stochastic equation that gives rise to the convective part in \eqref{eq:N_original}. This equation is
\be \label{eq:basic}
\frac{dx(t)}{dt}=  v(\vec{y}(t)) \,\, \Rightarrow\,\, x(t)= \int_0^t dt'v(\vec{y}(t'))
\ee
Intrinsic diffusive noise along the $x$-direction may be incorporated easily --- in particular, the mean will be unaffected while $D_{\sub{taylor}}$ will be additively augmented by $D$. We are considering the case $x(0)=0$ here. To proceed further, we introduce the Green's function/propagator $G(\vec{y},t|\vec{y}')$ -- the $t>0$ solution to \eqref{eq:n_original} when $n(\vec{y}, 0)=\d(\vec{y}-\vec{y}')$. It's most important use is encoded in the formula\cite{morse53}
\begin{equation}\label{eq:propagator-conc}
n(\vec{y},t) = \int_A dy' G(\vec{y},t-t'|\vec{y}')n(\vec{y}',t').
\end{equation}
for $t\geq t'$. In other words, it gives us the {\em fraction} of particles starting out from an arbitrary initial state $(\vec{y}',t')$ and surviving till a specific final state  $(\vec{y},t )$. 

To illustrate the technique of obtaining the moments of $x(t)$, we start with the first one -- the mean. It is obtained by averaging \eqref{eq:basic} over the particles that survive the absorbing walls till time $t$. Following \eqref{eq:propagator-conc}, the particle density at $(\vec{y}',t')$ is given by $\int G(\vec{y},t'|\vec{y}_i)n(\vec{y}_i,0)dy_i$. However, in the remaining time $(t-t')$ only a fraction $G(\vec{y}_f,t -t'|\vec{y})$ of these will survive till a final state $(\vec{y}_f,t )$. Thus, the number density of particles that pass through $(\vec{y}',t')$ and {\em also } survive till $t$ is
\be\label{eq:nu1} \nu(\vec{y},t'|t) = \int_A \!\!dy_f\!\! \int_A \!\!dy_i\,G(\vec{y}_f,t-t'|\vec{y})G(\vec{y},t'|\vec{y}_i)n(\vec{y}_i,0)
\ee
Multiplying this density by the corresponding displacement $v(\vec{y})dt'$, averaging over all possible intermediate positions and finally adding up all these averaged displacements, we find that
\be\label{eq:first_moment} \langle x(t) \rangle = \frac{1}{N(t)}\int_0^t dt' \!\!\int_A \!\!dy\,\, v(\vec{y})\nu(\vec{y},t'|t) \ee
where the (normalizing) denominator $N(t)$ is equal to the total number of particles that survive till $t$. Similarly, squaring \eqref{eq:basic} and averaging over the surviving particles one can show that the second moment is given by
\begin{widetext}
\bs\label{eq:second_moment}
\begin{align}
\langle (x(t))^2\rangle &=  \frac{2}{N(t)} \int_{0\leq t_1 \leq t_2 \leq t}\!\!\!\!\! dt_1 dt_2 \int_{A\otimes A}dy_1 dy_2 v(\vec{y}_1)v(\vec{y}_2) \nu(\vec{y}_1,t_1|\vec{y}_2,t_2|t)\\
\nu(\vec{y}_1,t_1|\vec{y}_2,t_2|t) &= \int_{A\otimes A} dy_i dy_f G(\vec{y}_f,t-t_2|\vec{y}_2)G(\vec{y}_2, t_2-t_1|\vec{y}_1)G(\vec{y}_1, t_1|\vec{y}_i)n(\vec{y}_i,0)
\end{align}
\es
\end{widetext}
The density of walkers who survive till $t$ after passing through $(\vec{y}_1,t_1)$ and $(\vec{y}_2,t_2)$ is given by $\nu(\vec{y}_1,t_1|\vec{y}_2,t_2|t)$, analogous to \eqref{eq:nu1}. All higher moments may be obtained by repeating the arguments leading to the above results.

The preceeding arguments thus provide the correct exposition of the stochastic technique to be used when absorption is also present. This will be the mainstay of the remaining results to be derived in this paper. We note that the formulation is exact in the framework we consider --- viz.\ the statistics obeyed by $\vec{y}$ is independent of $x$.

We can solve \eqref{eq:n_original} by expanding in terms of diffusion eigenmodes\cite{morse53}:
\be
\label{eq:n_sol} n(\vec{y}, t)=\sum_{k=0}^\infty n_k \psi_k(\vec{y})e^{-\l_k t}
\ee
where $n_0\neq 0$ (necessarily) and
\bs\label{eq:eigenfunctions}
\begin{align}
(D\,\nabla_{y}^2 + \lambda_k)\psi_k(\vec{y}) &= 0\qquad &\vec{y}\in A\\
(D\,\mbs{\hat{e}.\nabla}_{y} + \rho)\psi_k(\vec{y})  &= 0 &\vec{y}\in\partial A\\
(\lambda_0 < \lambda_1  < \lambda_2 &\ldots)\nn\\
\int_A \! dy \,\,\psi_j(\vec{y}) \psi_k(\vec{y}) &= \delta_{jk}
\end{align}
\es
The propagator is given by \cite{morse53}
\begin{equation}\label{eq:D}
G(\vec{y},t|\vec{y}') =\sum_{n=0}^\infty \psi_n(\vec{y})\psi_n(\vec{y}')e^{-\lambda_n t}
\end{equation}
The lowest eigenvalue $\l_0$ for this problem becomes non-zero for $\r \neq 0$ --- this is the main difference between the absorbing and non-absorbing cases. At long times all the higher modes decay with inverse mean lifetimes of
\begin{equation}
\l_n  \sim n^2/\t
\end{equation}
where $\t$ is the characteristic relaxation time-scale of the transverse diffusion process and scales as $\ell^2/D$ where $\ell$ is the characteristic length-scale of the cross-section $A$. In the long-time limit $t\gg\tau$ we can thus work with only the lowest mode in \eqref{eq:n_sol} and the errors will go as $\mc{O}(e^{-t/\t})$.

We consider this long time limit in what follows. We find that the moments can be conveniently evaluated in terms of simple `matrix elements'
\begin{equation}
v_{jk} =\int_A dy \,\,\psi_j(\vec{y})v(\vec{y})\psi_k(\vec{y})
\end{equation}
Using (\ref{eq:D}) and (\ref{eq:eigenfunctions}) in (\ref{eq:first_moment}) and keeping only the dominant terms as outlined in the previous paragraph we obtain the following expression for the first moment in the long time limit:
\begin{equation}
\label{eq:mean} \langle x(t) \rangle  \xrightarrow{t\gg 2\t} \overbrace{v_{00}}^{v_e} t + \mc{O}(e^{-t/\t})
\end{equation}
This result is very counter-intuitive since one na\"{\i}vely expects, at late enough times, the effective velocity $v_e$ to be the same as the asymptotic instantaneous mean velocity that is proportional to $\int_A v \,n \propto \int_A v \,\c_0$ ---  since $n(\vec{y}, t\rightarrow \infty)\sim \c_0(\vec{y})$. However, absorption at the walls modifies the probability that a particle starting out at $\vec{y}$ survives  after a long time (i.~e.\ $\gg \t$) from being $\vec{y}$-independent to being proportional to $\c_0(\vec{y})$ --- this provides the additional power of $\c_0$ in \eqref{eq:mean} (recall that $v_{00}=\int v\,\c_0^2$).

Analogously, this time using (\ref{eq:second_moment}) instead of (\ref{eq:first_moment}) above we calculate the variance in $x$ to be
\begin{equation}
\label{eq:variance} \langle(\delta x(t))^2\rangle \xrightarrow{t\gg3\tau} t\times 2\underbrace{\sum_{k=1}^\infty \frac{v_{k0}^2}{\lambda_k - \lambda_0}}_{D_{\sub{taylor}}} + \mc{O}(e^{-t/\t})
\end{equation}
while the third cumulant simplifies to:
\ba
\label{eq:thirdcumulant} \langle(\delta x(t))^3\rangle \xrightarrow{t\gg4\tau}  6t\sum_{j=1}^\infty\sum_{k=1}^\infty \frac{v_{0j}(v_{jk} - \d_{jk}v_{00})v_{k0}}{(\lambda_j - \lambda_0)(\lambda_k - \lambda_0)}
\end{align}
showing that the skewness, which is a measure of the deviation from Gaussianity, dies out with time:
\begin{equation}
\label{eq:skewness} \gamma = \frac{\langle(\delta x(t))^3\rangle}{\langle(\delta x(t))^2\rangle^{3/2}}\sim \sqrt{\frac{\tau}{t}}
\end{equation}
Using the same arguments, higher cumulants may be simplified to forms analogous to equations \eqref{eq:mean} or \eqref{eq:variance} in the long time limit. Because of the straightforward algebra involved, computer algebra systems (we used Mathematica) may be programmed to do this correctly and eliminate corrections $\mc{O}(e^{-t/\t})$.

We have analytically computed uptil the fifth cumulant this way and found that they are all proportional to time in the long-time limit. We assert that this holds to all orders. This implies that the p.d.f of the longitudinal displacement tends to a Gaussian with deviations that die as $1/t$ --- similar to the case of reflecting boundary conditions\cite{aris56}. A similar behaviour is also exhibited by drifting random walkers on a lattice\cite{haldane40}.

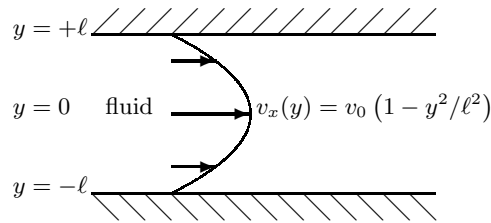
\begin{figure}[ht]
\begin{picture}(150,100)(20,0)
\thicklines \put(20,20){\line(1,0){130}}
\put(20,80){\line(1,0){130}} \thinlines
\qbezier(50,80)(110,50)(50,20)
\multiput(20,20)(10,0){13}{\line(1,-1){10}}
\multiput(20,80)(10,0){13}{\line(1,1){10}}
\put(82,50){$v_x(y)=v_0\le(1-y^2/\ell^2\ri)$}
\put(25,50){fluid} \put(-10,80){$y=+\ell$} \put(-10,20){$y=-\ell$}
\put(-10,50){$y=0$} \thicklines \put(50,50){\vector(1,0){30}}
\multiput(50,30)(0,40){2}{\vector(1,0){17}}
\end{picture}
\caption{Taylor dispersion between parallel plates
\label{fig:plates}}
\end{figure}

Finally let us compute a specific example -- Taylor dispersion of particles suspended in fluid flowing unidirectionally between two infinite parallel plates (see fig.~\ref{fig:plates}). Absorption at the walls is characterised by the dimensionless parameter $\a=\r \ell/D$ ($\r$ is defined in \eqref{eq:bc_original}). For convenience, we have used the dimensionless coordinate $\y=y/\ell$ below. The eigenfunctions \eqref{eq:eigenfunctions} for this case turn out to be:
\bs \ba \c_n(\y) &= \sqrt{\frac{\a^2 + \f_n^2}{(\a+\a^2+\f_n^2)}}\cos\bigg(\f_n\y+\frac{n\p}{2}\bigg) \\
\frac{\a}{\f_n}&= \tan\bigg(\f_n +\frac{n\p}{2}\bigg),\qquad n=0,1,\ldots
\end{align}
\es
We obtained, for $\a\ll 1$
\bs \ba
\f_n  &= \sqrt{\a}\le(1-\frac{\a}{6} + \frac{11\a^2}{360} + \mc{O}(\a^3)\ri), &n=&0\\
&= \frac{n\p}{2} + \frac{2\a}{n\p}-\frac{2}{n\p}\le(\frac{2\a}{n\p}\ri)^2 +\mc{O}\le(\frac{2\a}{n\p}\ri)^3, &n=&1\ldots
\end{align}
\es
For $\a\rightarrow\infty$,
\be \f_n = \frac{(n+1)\p}{2},\qquad n=0,1,\ldots \ee
Finally, the previously introduced eigenvalues $\l_n$ (see (\ref{eq:eigenfunctions})) are obtained from the relation:
\be \l_n  = D \frac{\f_n^2}{\ell^2} \ee

Using our formul\ae\  \eqref{eq:mean} and \eqref{eq:variance} we have calculated the effective velocity and the Taylor dispersion constant to be:
\bs\label{eq:parallel_plate_1}
\ba
v_e &= \frac{2v_0}{3}\le(1+\frac{2}{15}\a + \mc{O}(\a^2)\ri),\a \ll 1\nn\\
&= \frac{2v_0}{3}\le(1+\frac{3}{\p^2}\ri),\quad\a\rightarrow\infty \\
D_{\sub{taylor}}&= \frac{8 v_0^2 \ell^2}{945 D}\le(1-\frac{4}{15}\a+\mc{O}(\a^2)\ri), \a\ll 1 \nn\\
&= \mbox{ $14\%$ of value at $\a=0$, when $\a\rightarrow \infty$}
\end{align}
\es
When there is no absorption ($\a=0$), $v_e$ is equal to the asymptotic mean instantaneous velocity of the particles since $\c_0$ is constant. However, since the slower particles near the walls are also the ones that are preferentially absorbed, $v_e$ increases with $\a$ as only surviving particles contribute to the moments. Similarly, since more particles at the walls get removed for increasing $\a$, the dispersion in the convection field as seen by the particles goes down and so does $D_{\sub{taylor}}$ which arises due to it. Our results \eqref{eq:parallel_plate_1} agree with those derived in \cite{lungu82} by a more tedious method. One may reproduce (\ref{eq:variance}) from \cite{lungu82} with some effort. However, as already noted before, it is very difficult to obtain simple formul\ae\  for the higher moments using the method presented there.

Finally, the skewness, which is a measure of the deviation from gaussianity, is found to be
\ba\label{eq:parallel_plate_2}
\g &= -0.186\sqrt{\frac{\ell^2}{D t}}\le(1 + 1.7 \a + \mc{O}(\a^2)\ri),\a \ll 1\nn\\
&= -0.635 \sqrt{\frac{\ell^2}{D t}},\quad\a\rightarrow\infty
\end{align}
showing that it is negative and enhanced due to absorption at the walls. The zero-absorption case of this result \eqref{eq:parallel_plate_2} has been independently verified using the method presented in \cite{aris56}.

In conclusion, we have a picturesque and computationally simple solution to a problem of general interest. For example, we may now estimate the error in measuring $D$ using Taylor dispersion\cite{cussler97}. The value of $\l_0$ and consequently $\a$ may be obtained from the observed exponential decay of the particle numbers and subsequently used in the corrected formula (\ref{eq:variance}) to estimate $D$ from the experimentally-observed variance. Also note that the moments in the case of an arbitrary cross-section may be computed numerically to the necessary precision by involving a finite number of eigenmodes. This is possible because terms involving higher eigenmodes get progressively less important as may be easily seen from the forms of \eqref{eq:mean} and \eqref{eq:variance}.

\begin{acknowledgements}
We would like to thank Prof.\ B.\ I.\ Halperin for his insightful comments and Srividya I.\ Biswas for stimulating discussions. RRB would also like to thank Schlumberger-Doll Research for supporting this work.
\end{acknowledgements}


\end{document}